\documentclass[preprint,final,aps,byrevtex]{revtex4}
\usepackage{amsfonts}
\usepackage{amsmath}
\usepackage{amssymb}
\usepackage{graphicx}
\usepackage{booktabs}
\usepackage{multirow}

\begin{document}

\title{Electronic and magnetic properties of bilayer graphene with
intercalated adsorption atoms C, N and O}
\author{S. J. Gong$^{1}$\cite{G}, W. Sheng$^{2}$, Z. Q. Yang$^{2}$, J. H. Chu%
$^{1,3}$} \affiliation{1. National Laboratory for Infrared Physics,
Shanghai Institute of Technical Physics, Chinese Academy of
Sciences, Shanghai 200083, China} \affiliation{2. National
Laboratory for Surface Physics, Fudan University, Shanghai 200433,
China.} \affiliation{3. Key Laboratory of Polar Materials and
Devices, Ministry of Education, East China Normal University,
Shanghai 200241, China}

\begin{abstract}
We present an $ab$-$initio$ density function theory to investigate the
electronic and magnetic structures of the bilayer graphene with intercalated
atoms C, N, and O. The intercalated atom although initially positioned at
the middle site of the bilayer interval will finally be adsorbed to one
graphene layer. Both N and O atoms favor the bridge site (i.e. above the
carbon-carbon bonding of the lower graphene layer), while the C atom prefers
the hollow site (i.e. just above a carbon atom of the lower graphene layer
and simultaneously below the center of a carbon hexagon of the upper layer).
Concerning the magnetic property, both C and N adatoms can induce itinerant
Stoner magnetism by introducing extended or quasilocalized states around the
Fermi level. Full spin polarization can be obtained in N-intercalated system
and the magnetic moment mainly focuses on the N atom. In C-intercalated
system, both the foreign C atom and some carbon atoms of the bilayer
graphene are induced to be spin-polarized. N and O atoms can easily get
electrons from carbon atoms of bilayer graphene, which leads to Fermi level
shifting downward to valence band and thus producing the metallic behavior
in bilayer graphene.

%We also provide some comparison about the adsorption effect between
%the monolayer and bilayer graphene.

%Three possible adsorption sites have been
%considered for each adatom. One of the adsorption sites is above the
%center of a carbon hexagon of the lower graphene layer (called the
%Hollowsite), the second is directly on top of a carbon atom of the
%lower graphene layer and simultaneously right below a carbon atom of
%the upper layer (called the Top site), and the third is above the
%carbon carbon bond of the lower graphene layer (called the Bridge site).

\textbf{\medskip }PACS Numbers: 71.15.Mb, 73.43.Cd, 81.05. Uw
\end{abstract}

\maketitle

%\thanks{To whomcorrespondence should be addressed} \email[Email address:]{zyang@fudan.edu.cn}

\textit{\ }\textbf{Introduction:} Since the discovery of monolayer graphene
in the year 2004 \cite{Novo}, this two-dimensional material remains in the
focus of active research motivated by its novel physical properties and a
promising potential for applications\cite{Tan,Gusy,Wang}. Experimental
groups have enabled preparation and study of systems with one or a small
number of graphene layers \cite{Geim}. Bilayer graphene, which is made of
two stacked graphene layers, is considered to be particularly importance for
electronics applications because of its special band structure \cite{Min}.
Coupling of the two monolayer graphene sheets in the usual A-B stacking of
bilayer graphene yields pairs of hyperbolically dispersing valence and
conduction bands that are split from one another by the interlayer
interaction\cite{Mcca}. The band gap of the bilayer can be easily opened
when a difference between the electrostatic potential of the two layer is
introduced, either by chemical doping or by applying gate voltage \cite%
{Zhong,Kuzm,Nils, Nico, Bouk,Ohta,Zhou,Kosh,Yu}, which makes graphene
channels have a high resistance for the OFF state. Angle-resolved
photoemission (ARPES) measurements indicate such a gap in potassium doped
bilayer graphene epitaxially grown on SiC \cite{Ohta}, and infrared
spectroscopy measurements also detect the similar gap in the
electrostatically gated bilayer graphene \cite{Kuzm,Nils,Nico}.

%Even very dilute impurity in graphene may induce obvious changes
%from semiconductor to metal or insulator, nonmagnetic to magnetic, etc.%
%ÔõôÒýµ½ÔÓÖÊÉÏÀ´£¿

%Compared with monolayer graphene, investigations on doping effect in
%bilayer graphene are rare, which should be attached much more
%importance to.

In this work, we carried out first-principle calculations and theoretical
analysis to explore the electronic and magnetic properties of bilayer
graphene with intercalated atoms C, N and O. The calculations were performed
using the projector augmented wave (PAW) formalism of density functional
theory (DFT) as implemented in the Vienna $ab$ initio simulation package
(VASP) \cite{Kres}. Because generalized gradient approximation (GGA) \cite%
{Perd} gives essentially no bonding between graphene planes and leads to
excessively large values of bilayer distance \cite{Bouk}, we performed the
calculations within the localized density approximation (LDA). We find that
LDA gives rise to a bilayer distance of 3.34 $\mathring{A}$, in good
agreement with the experiment value. Some previous investigations used LDA
to optimize the structure to get a reasonable bilayer distance, and
sequently used GGA to calculate the electronic structure \cite{Zhong}. To
keep consistent, we use LDA in all our calculations for the investigated
system. An energy cutoff of 400 eV for plane-wave expansion of the PAWs is
used. The model system here consists of a $4\times 4$ supercell with a
foreign atom intercalated between the two coupling graphene layers. The
supercell parameters are set to be the same as $a=b=9.84$ $\mathring{A}$ in
the $xy$ plane ($a$ and $b$ indicate the crystal lattice constants). The
Brillouin zone is sampled using a $11\times 11\times 1$ $\Gamma $ centered
k-point grid. For geometry optimization, all the internal coordinates are
relaxed until the Hellmann-Feynman forces are less than 0.01 eV/$\mathring{A}%
.$ The vacuum thickness along the z axis is 16 $\mathring{A}$ to avoid the
interaction between graphene layers of adjacent\ supercells.

\bigskip We considered the $\widetilde{\mathbf{A}}$-B Bernal stacking
structure for bilayer graphene. Top-view for the bilayer graphene is shown
in Fig. 1(a), in which violet color is for the lower layer, and gray color
is for the upper layer. Seen from the top-view, `$\widetilde{\text{A}}$'
position in the lower layer coincides with `B' position in the upper layer,
`A' position in the upper layer is exactly the center of the hexagon of the
lower graphene layer from the top-view, and `$\widetilde{\text{B}}$'
position in the lower layer is right the center of the hexagon of the upper
graphene layer. The foreign atom is intercalated between the upper and lower
graphene layers and is initially positioned in the middle position of the
bilayer distance. Three kinds of initial positions are considered for each
intercalated atom: bridge, top and hollow positions, which are noted in the
following by using subscript $^{\text{\textbf{\textquotedblleft }}}$%
Bri\textquotedblright , \textquotedblleft Top\textquotedblright , and
\textquotedblleft Hol\textquotedblright\ when needed, for example, we use N$%
_{\text{Bri}}$\ to indicates the bilayer system with N atom positioned at
the bridge site. Bridge site is above the carbon-carbon bonding in the lower
graphene layer, top site indicates the middle position between the
coinciding positions `$\widetilde{\text{A}}$' and `B', and hollow site is
just above a carbon atom of the lower graphene layer and simultaneously
below the center of a carbon hexagon of the upper layer. The structure
relaxation calculations show that, if the intercalated atom is initially
positioned at the bridge or hollow site, it will finally be adsorbed to one
graphene layer and away from the other layer. While for the top site, the
foreign atom will keep being located in\textbf{\ t}he middle position of the
bilayer interval.

The binding energy is defined as: $dE=E_{\text{graphene}}+E_{\text{atom}}-E_{%
\text{total}},$ where $E_{\text{graphene}}$ is the energy of the clean
bilayer graphene, $E_{\text{atom}}$ stands for the energy of the single
foreign atom, and $E_{\text{total}}$ is the total energy of the bilayer
graphene with the intercalated atom. The binding energy, the distance
between the foreign atom and its nearest C atom, and the interlayer distance
of the doped bilayer graphene are illustrated in Table I. Comparing the data
in Table I, we find both N and O atoms favor the bridge site, while the C
atom prefers the hollow site. No intercalated atom is stable at the top
site. At the top position, the relaxation calculations show that nearest C-N
bonding is about 1.83 $\mathring{A}$, and C-O bonding is about 1.74 $%
\mathring{A}$, which is much larger than the typical lengths of C-N (1.47 $%
\mathring{A}$)and C-O (1.42 $\mathring{A}$) \cite{Bond}, implying the
physisorption rather than the chemisorption. For the hollow site, O atom is
adsorbed to the layer which provides the nearest C atom, and away from the
hexagonal cental of the other graphene layer. O atom and the nearest C atom
form the C=O bonding with the length of 1.39 $\mathring{A}$, smaller than
the the length of C-O bonding in the O$_{\text{Bri}}$ structure (1.44 $%
\mathring{A}$). The calculation shows N$_{\text{Hol}}$ structure doesn't
exist because of the negative binding energy. Concerning the bilayer
distance, with the foreign atom intercalated in the bilayer space, the
distance is enlarged for all the investigated systems.

Figure 1 displays the ground states for C, N and O configurations, where
Fig. 1(b), (c) and (d) are the C$_{\text{Hol}}$, N$_{\text{Bri}}$, and O$_{%
\text{Bri}}$ structures, respectively. In the C$_{\text{Hol}}$ system, the
optimized interlayer distance is about 3.39 $\mathring{A}$, which is a
little larger than the value of 3.34 $\mathring{A}$ for the pure bilayer
graphene. Calculations show that C-C bonding in the upper layer nearly keeps
unchanged. The carbon atom which is right below the foreign C is pushed
down, forming a \textquotedblleft dumbbell\textquotedblright\ at the saddle
point. The foreign C bonds with the adjacent three carbon atoms with bonding
length being 1.54 $\mathring{A}$, which is a standard length for $sp^{3}$
hybridization \cite{Bond}. In N$_{\text{Bri}}$ system, the bilayer distance
is enlarged to the value of 3.87 $\mathring{A}$, and the length of C-N is
1.43 $\mathring{A}$, indicating the chemically adsorption not the physically
adsorption. The two carbon atoms bonded with N atom are drawn out of the
graphene layer, and the C-C bonding is 1.55 $\mathring{A}$, which implies $%
sp^{3}$ hybridization of the two carbon atoms. In O$_{\text{Bri}}$ system,
bilayer distance is about 3.76 $\mathring{A}$, length of C-O bonding is 1.44
$\mathring{A}$, and the two carbon atoms bonded with O atom are also drawn
out of the graphene layer with the C-C bonding is 1.51 $\mathring{A}$.
Obviously, N and O atoms have the similar ground state structure.
%Ë«¼üË«¼üË«¼ü

%For N and O atom adsorption, the previous investigation showed that
%in monolayer graphene, their stable sites are both at the Bridge
%site \cite{Jiang}. In bilayer graphene, our calculation shows that
%their stable positions are still at the Bridge site. For the stable
%Hole-C system, the additional C atom pushes down one of the
%underlying C atom, and forms a "dumbell" at the saddle point. For
%the ground state of a C adatom on the monolayer graphene surface,
%the stable site is the Bridge site \cite{Ma}. We will explain the
%difference of C adsorption in bilayer and monolayer graphene in the
%following. In the Hole-C system, the adatom C obviously bonds with
%the adjacent carbon atom to form sp3 hybridization with C-C bonding
%length being 1.54 $\dot{A}$. In Bridge-N and Bridge-C system, The
%carbon atoms bonded with adsorption atom also tries to form sp3
%hybridization, with C-C bonding being 1.55 and 1.51 $\dot{A}$,
%respectively. Length of C-N and C-O are 1.43 and 1.44 $\dot{A}$,
%respectively, indicating the chemically adsorption, not the
%physically adsorption.

%\textbf{Results and discussion:}

Side-view of the charge distributions for C$_{\text{Bri}}$, N$_{\text{Bri}}$%
, and O$_{\text{Bri}}$ systems are displayed in Fig. 2. Although C$_{\text{%
Bri}}$ structure is not the ground state for C-intercalated system, we show
its charge contour together with those of N$_{\text{Bri}}$ and O$_{\text{Bri}%
}$ systems, to provide a clear comparison. We know O atom is lack of two
electrons, when it is adsorbed in the bilayer graphene, it strongly
interacts with its adjacent carbon atoms and get electrons from them (see
the upper panel of Fig. 2). N atom lacks three electrons, it also gets
electrons from its adjacent carbon atoms (see the middle panel of Fig.2). C
atom lacks four electrons to get saturated, and it tends to share electrons
with its adjacent carbon atoms in bilayer graphene (see the lower panel of
Fig. 2). All these three atoms show strong interaction with lower graphene
layer. Concerning their interactions with the upper graphene layer, judging
from Fig. 2, C atom has the strongest interaction with the upper layer, N is
the second, and O is the third. We compare the stable positions for C, N and
O atoms in monolayer and bilayer graphene, as illustrated in table II. It is
clear that N atom has the same stable position in both monolayer and bilayer
graphene, so does the O atom. C adatom is stable at the bridge site in
monolayer graphene, and Hollow site in bilayer graphene. We believe the
different stable positions for C atom results from its interaction with the
upper graphene layer.

The C$_{\text{Hol}}$ structure is favored as the ground state for C atom
intercalated in the bilayer graphene. The spin-resolved band structures and
density of states (DOS) for C$_{\text{Hol}}$ system are shown in Fig. 3.
Fig. 3(a) and (b) are the band structures for the majority spin and minority
spin, respectively. It is clearly seen that impurity bands for the majority
and minority spin components lie, respectively, lower and higher than the
Fermi level. In the spin-resolved total DOS (see Fig. 3(c)), two narrow
peaks at the opposite side of the Fermi level are observed, indicating the
itinerant magnetism triggered by the intercalated C atom. In the following,
we will find the itinerant magnetism comes from not only the foreign C atom,
but also the carbon atoms of the bilayer graphene. In the orbital-resolved
PDOS of the foreign C atom, $s$ state and $p$ state have peaks in the same
energy range, which is indicative of the $sp^{3}$ hybridization. The C-C
bonding between the foreign C and the nearest carbon atoms with the value of
1.54 $\mathring{A}$, is also a proof of the $sp^{3}$ hybridization, which
has been pointed in the previous analysis. When the spin degree of freedom
is neglected, our calculation from first principle predicts a twofold
degenerate peak at the Fermi level (Fig. 3(e)), but the spin unpolarized
state is not the ground state. Including the spin degree of freedom, the
balance between the majority and minority spin components will be destroyed.
The spin density distribution of the lower graphene layer of the C$_{\text{%
Hol}}$ system are shown in Fig. 3(f). Both the foreign C atom and the pushed
down carbon atom are magnetic, yet we cannot see their magnetic moment
distribution in Fig. 3(f), because they are not in the lower graphene plane.
Seen from the top-view, these two carbon atoms occupy coinciding positions,
which are noted in Fig. 3(f) by the letter `C'. The three nearest carbon
atoms bonded with the foreign C atom are nearly nonmagnetic and the
next-near-neighbors are magnetic. On the whole, the total magnetic moment of
the C$_{\text{Hol}}$ system is about 1.32 $\mu B$, and the magnetic moment
distributions show threefold symmetry, which is similar with the hydrogen
adsorption on the graphene plane \cite{Yazy}.

%For monolayer graphene with adsorption atom C, our calculation shows
%that its stable site is at the Bridge site. From the charge contour
%of Fig. 2, it can be seen that adsorption atom C has stronger
%interaction with the upper layer, compared with N and O atom. The
%magnetic moment of the Hole-C structure are distributed.

\bigskip Spin-resolved band structure and density of states for N$_{\text{Bri%
}}$ system are shown in Fig. 4, where Fig. 4(a) and (b) show the band
structure for majority and minority spin, respectively, and Fig. 4(c) and
(d) display the total DOS of N$_{\text{Bri}}$ system and PDOS of N atom,
respectively. In the total DOS, very narrow and sharp peak at the Fermi
level is observed. We draw the PDOS of N atom and find the peak arises from
from the N adsorption and the N atom is nearly 100\% spin polarized. In the
band structure, we find the very localized states at the Fermi level in the
minority spin band structure. Such quasilocalized states give rise to the
strong Stoner magnetism with magnetic moment of 0.65 $\mu $B located at the
N atom. In the PDOS of N atom, between the energy -0.5 eV and -1 eV, another
narrow peak is obtained. Both in the majority and minority spin band
structures, the corresponding flat impurity bands in the energy range from
-0.5 eV to -1 eV are observed. In addition, both in the majority and
minority band structures, the characteristic conical point at K point still
can be clearly identified, implying the bilayer graphene is not strongly
purturbed by the N atom. The critical difference is that, in freestanding
graphene the Fermi level coincides with the conical point, while the Fermi
level obtained in N-intercalated system is shifted downward and becomes
below the conic point. A shift downwards (upwards) means the holes
(electrons) are donated by the adsorption atom. The manganese doping
(electron donor) results in the upward shift of the Fermi level, reported by
previous investigations \cite{Zhong}.

%For the O adsorption system, no polarized electrons are detected, resulting
%in no magnetic moment. For N and C adsorption system, at the Bridge site,
%they cannot get enough electrons from adjacient C atoms to to get saturated.
%The unpaired electron in N adsorption system is polarized, and the
%magnetization mostly focuses on the N atom. From the partial dos of N atom,
%we can see that N atom is 100\% spin-polarized around its Fermi level.
%Monolayer graphene with N adsorption atom has been investigated previously,
%and the reported magnetization is about 0.84 $\mu B$ with the generalized
%gradient approximation of Perdew et al. for exchange correlation density
%function \cite{Novo}. In bilayer graphene, the structure relaxation should
%be conducted with localized density function because of the interlayer
%coupling is van der Waals force\cite{Novo}. To keep consistent, structure
%oprimization, electronic structure, and magnetic properties all are
%calculated within the localized density function approximation (LDA) in our
%work. The magnetization of N adsorbed monolayer graphene system is about
%0.68 $\mu $B with LDA calculation, weaker than that within GGA calculations.
%For bilayer graphene with N adsorption atom, the magnetization is about 0.65
%$\mu $B, which is smaller than that in monolayer graphene. In bilayer
%graphene, the N atom can interact with the C atoms in the upper layer, which
%reduces the unsaturated electrons of N atom. For Bridge-C structure, the
%adsorption C atom is also polarized, and the magnetization is about 1.32 $%
%\mu B$ .

\bigskip

Band structure and density of states of O$_{\text{Bri}}$ system are
displayed in Fig. 5, where Fig. 5(a), (b) and (c) are the band structure,
total density of states, and partial density of states of O atom,
respectively. The calculation shows the O$_{\text{Bri}}$ system is
nonmagnetic. From the above analysis about N$_{\text{Bri}}$ and C$_{\text{Hol%
}}$ systems, we see the impurity states localized around the Fermi level
play an important role in the magnetic properties, and the intercalated atom
itself is spin-polarized. In O$_{\text{Bri}}$ system, we do not get such
localized states around the Fermi level, and the O atom is nonmagnetic. In
the PDOS of O atom, we find the peak nearest to the Fermi level is around
the energy of 0.75 eV. From the band structure, we can also see that the
impurity state nearest to the Fermi level is at the energy of 0.75 eV. Like N%
$_{\text{Bri}}$ system, the O$_{\text{Bri}}$ system also undergoes a change
from semimetal to metal, because O atom get electrons form graphene, thus
the Fermi level shifts downward.

%We finally give the comparison of the magnetic moment in monolayer
%and bilayer graphene. For N and O, their stable position is the
%Bridge site, both for monolayer graphene and bilayer graphene. For
%adsorption atom C, in the monolayer graphene, its stable position is
%the Bridge site, while for bilayer graphene, its stable position is
%the Hollowsite. In table-II, the stable position for N, C and O atom
%are listed. This can be understood form the interaction between the
%adsorption atom and the graphene layers. For adatom C, it has
%stronger interaction with the upper layer graphene compared with N
%and O, which leads to the difference situation between the monolayer
%and bilayer graphene. For the magnetic properties, generally,
%magnetic moment is smaller than that in monolayer graphene. We
%ascribe this phenomenon to the weak bonding between the adatom and
%the upper layer graphene, which decreases the number of the unpaired
%electrons.

%\begin{table}[tbph]
%\begin{center}
%\includegraphics*[width=10cm]{table3.eps}
%\end{center}
%\caption{Comparison of the magnetic moment in monolayer and bilayer graphene
%layer. }
%\end{table}

In summary, we calculated the structure, electronic and magnetic properties
of the bilayer graphene with intercalated atoms C, N, and O by the $ab$-$%
initio$ density function theory. Impurities even at very low density may
bring fruitful physical properties for bilayer graphene system. Structure
relaxation shows that the intercalated atom although initially positioned at
the middle position of the bilayer distance, will finally be adsorbed to one
graphene layer and away from the other layer. The bilayer distance is
enlarged for all the investigated adsorption systems. Both N and O atoms are
stable at the bridge site, while C atom is stable at the hollow site.
Concerning the magnetic property, O-intercalated system is nonmagnetic,
while N and C atoms can induce stoner magnetism by introducing extended or
quasilocalized states around the Fermi level. Nearly 100\% spin polarization
is obtained in N-intercalated system and the magnetic moment focuses on the
N atom. For C-intercalated system, the magnetic moment distributes on the
foreign C atom and certain carbon atoms of the bilayer graphene. Both in
N-intercalated and O-intercalated systems, the Fermi level is obviously
shift downward, inducing the metallic behavior of the bilayer graphene.

\textbf{Acknowledgements:} This work was supported by the the
Ministry of Sciences and Technology through the 973-Project (No.
2007CB924901), the National Natural Science Foundation of China
(Grant Nos. 60221502 and 1067027), the Grand Foundation of Shanghai
Science and Technology (05DJ14003).

\newpage

\section{Figure Captions}

\begin{description}
\item[Fig. 1] (a) $\widetilde{\text{A}}$-B Bernal stacking
structure for bilayer graphene. (b) C$_{\text{Hol}}$ system, where
the atom in violet is the foreign C atom. (c) N$_{\text{Bri}}$
system, where the atom in blue is the foreign N atom. (d)
O$_{\text{Bri}}$ system, where the atom in red is the foreign O
atom.

\item[Fig. 2] Side-view of charge contours for C$_{\text{Bri}}$, N$_{\text{Bri}}$
and O$_{\text{Bri}}$ systems, respectively.

\item[Fig. 3] Electronic and magnetic structures of the C$_{\text{Hol}}$ system.
(a) Band structure of majority spin, (b) band structure of minority
spin, (c) total density of states, (d)the orbital-resolved DOS for
the foreign C atom, (e) density of states for spin unpolarized
C$_{\text{Hol}}$ system, (f) distribution of the spin density on the
lower graphene layer..

\item[Fig. 4] Spin-resolved band structure and density of states of N$_{\text{Bri}%
}$ system. (a) Band structure of the majority spin; (b) band
structure of the minority spin, (c) total density of states, (d)
partial density of states of N atom.

\item[Fig. 5] Band structure and density of states of O$_{\text{Bri}}$ system.
(a) Band structure, (b) total density of states, (c) partial density
of states of O atom.

\item[Table I] The binding energy ($dE$), the length of bonding between the
adsorption atom and its nearest C atom ($a_{\text{C-atom}}$), and
the bilayer distance of the adsorption system ($dis.$). The stable
position for each intercalated atom is noted by box.

\item[Table II] The stable positions for adsorption atoms C, N, and O in monolayer
and bilayer graphene.

\end{description}

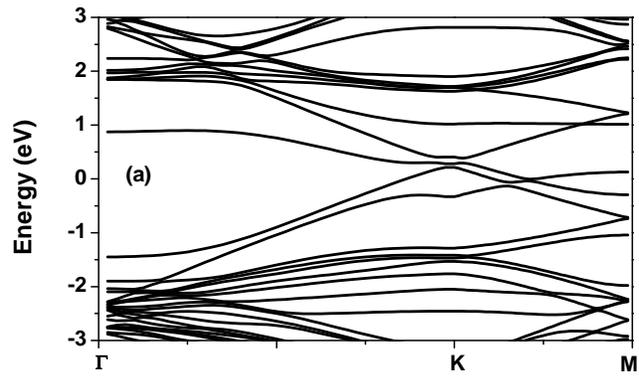
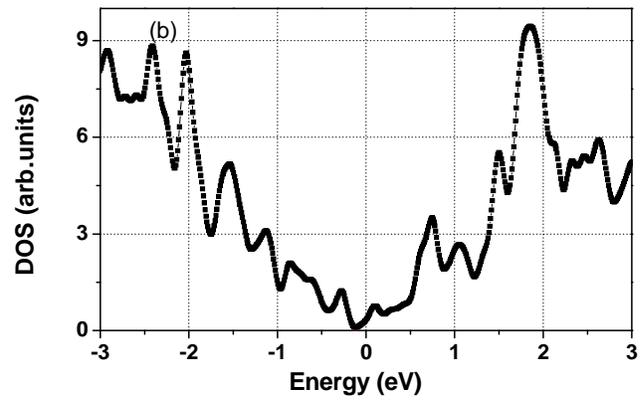
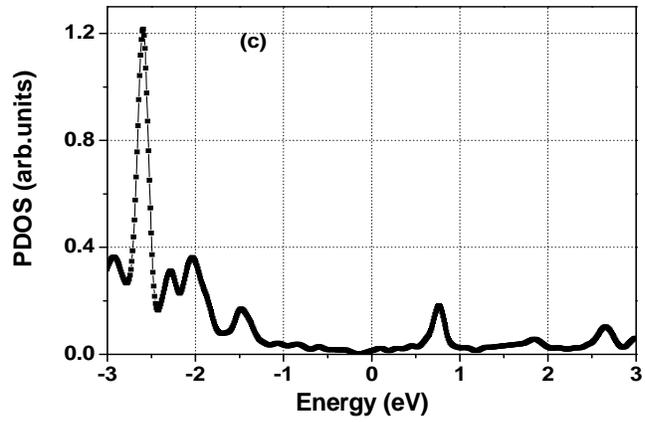
\end{document}